%% file: bomben_tipp11.tex
\documentclass[3p,times,procedia]{elsarticle}

\usepackage{ecrc}

\volume{00}

\firstpage{1}

\journalname{Physics Procedia}

\runauth{}

\jid{phpro}

\jnltitlelogo{Physics Procedia}

\CopyrightLine{2011}{Published by Elsevier Ltd.}

\usepackage{amssymb}

\usepackage[figuresright]{rotating}

\begin{document}

\begin{frontmatter}



\dochead{}

\title{Recent progress of the ATLAS Planar Pixel
Sensor R\&D Project}


\author[LPNHE]{M.~Bomben\corref{cor1}\fnref{ref1}}
\fntext[ref1]{on behalf of the ATLAS Planar Pixel
Sensor R\&D Project  \\ \indent \indent {\tt https://twiki.cern.ch/twiki/bin/view/Atlas/PlanarPixelUpgrade}}
\cortext[cor1]{corresponding author}

\address[LPNHE]{        Laboratoire de Physique Nucleaire et de Hautes
 \'Energies (LPNHE)
 \\ Tour 12-22, 4, place Jussieu, FR-75252 Paris Cedex 05}
  \ead{marco.bomben@lpnhe.in2p3.fr}

\begin{abstract}
\input{abstract}
\end{abstract}


\end{frontmatter}



\input{intro}

\input{project}

\input{p_bulk}

\input{n_bulk}

\input{thin}

\input{slim}

\input{active}

\input{simu}

\input{outl}





\bibliographystyle{elsarticle-num}
\bibliography{bomben_tipp11}








\end{document}

%% file: abstract.tex
The foreseen luminosity upgrade for the LHC (a factor of 5-10 more in peak luminosity by 2021) 
poses serious constraints on the technology for the ATLAS tracker in this High Luminosity 
era (HL-LHC). In fact, such luminosity increase 
leads to increased occupancy and radiation damage of the  tracking detectors.

To investigate the suitability of pixel sensors using the proven planar technology 
    for the upgraded tracker, the ATLAS
Planar Pixel Sensor R\&D Project was established comprising 17 institutes and more than 80 
scientists. 
    Main areas of research are the performance of planar pixel sensors at highest fluences, 
    the exploration of possibilities for cost reduction to enable the instrumentation of large areas, 
    the achievement of slim or active edge designs to provide low geometric inefficiencies without
the need for shingling of modules and the investigation of the operation of highly irradiated 
    sensors at low thresholds to increase the efficiency.

In the following I will present results from the group, concerning mainly irradiated-devices 
performance, together with studies for new sensors, including detailed simulations.

%% file: intro.tex
\section{Introduction}
\label{sec:intro}
ATLAS~\cite{ATLAS}  is a general purpose detector for the study of primarily proton-proton
collisions at the LHC~\cite{LHC}.

The ATLAS Inner Detector~\cite{AtlasID1,AtlasID2} provides charged-particle tracking 
with high
efficiency over the pseudorapidity range $|\eta|< 2.5$. 
The pixel detector system~\cite{AtlasPixels} is the innermost
element of the Inner Detector. The pixel detector contains approximately 80 million channels
and provides pattern recognition capability in order to meet the track reconstruction requirements
of ATLAS at the full luminosity of the LHC of $\mathcal L ={\rm 10^{34} cm^{-2}s^{-1}}$.

Consisting of three barrel layers (at radii between 50.5mm and 122.5mm),
and six discs, the pixel detector counts a total of 1744 pixel modules, which are mounted allowing
for a three-hit track-reconstruction of charged secondary-particles. 
Each module contains a 250 $\mu$m thick 
n-in-n pixel sensor of 62.6 $\times$ 18.6mm$^2$ with pixel cells of
50$\times$400 $\mu m^2$. Connected to each sensor are 16  7.4$\times$11.0mm$^2$
ATLAS FE-I3~\cite{FEI3} readout chips with a total of 46080 channels.
Both the sensors as well as the electronics of the present
ATLAS pixel modules were specified to work up to a fluence
of ${\rm 10^{15} n_{eq}/cm^2}$ and an ionising dose of 50 MRad.

\begin{figure}[!htb]
\centering
\includegraphics[height=0.35\textheight]{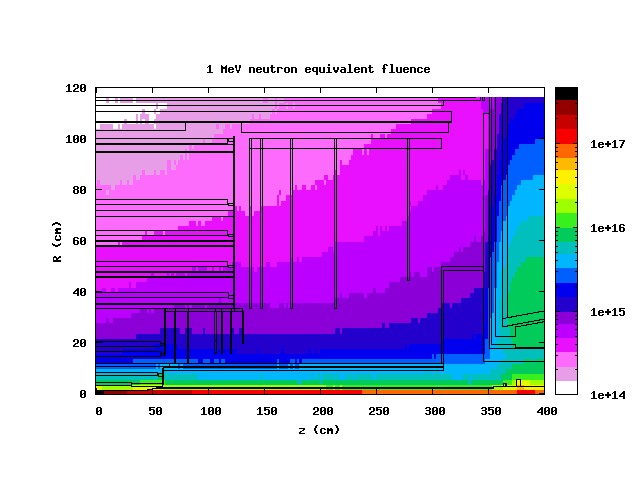}
\centering
\caption{\label{fig:irrad}Fluences in $n_{eq}/cm^{2}$ expected for
a phase II integrated luminosity of 3000 fb$^-1$. A safety factor of 2 is applied.}
\end{figure}

While the nominal luminosity of the present LHC accelerator
is ${\rm 10^{34} cm^{-2}s^{-1}}$, an upgrade to increase the luminosity
by a factor of ten (five with luminosity leveling) is planned to be carried out in a two phase
process~\cite{SLHC}.
After a first shutdown, foreseen for 2017, the {\it Phase 1} of LHC will start, with the goal 
luminosity of ${\rm (2-3) \times 10^{34} cm^2s^{-1}}$. 
A second shutdown will take place after 2020; then the {\it Phase 2} will start, and the expected 
luminosity is ${\rm 5 \times 10^{34} cm^{-2}s^{-1}}$ with luminosity levelling.
By 2030 a total integrated luminosity of ${\rm O(3000 fb^{-1})}$ is envisaged. 
Based on this scenario the innermost layer of the ATLAS pixel system will have to sustain 
fluences above $1\times {\rm 10^{16}\; n_{eq}/cm^2}$~\cite{fluences}; see also figure~\ref{fig:irrad}.

The main effects on the pixel sensors from these large fluences are an increase of the leakage 
current and a reduction in the charge collection efficiency. The final goal is to retain a good 
hit-efficiency up to the final integrated luminosity.

Another high-luminosity related effect is the increase of the occupancy of the detector. 
A natural option is to reduce the size of the elementary pixel cell; at the same time the 
spatial resolution will improve. The reduced size of the elementary cell is just one of the 
features of the new read-out chip for the pixel system, the FE-I4~\cite{FEI4} chip; 
the total surface of the FE-I4 chip is 20.0 $\times$ 18.6 mm$^2$.
Moreover, replacing the strips of the Inner Detector with pixel detectors might be an option to 
handle the occupancies foreseen at the HL-LHC.

Hence, in view of a possible pixel system replacement in 2017; 
and then, maybe, for a whole new tracking system after 2020, a new Pixel System is 
unders study.  
The new pixel sensors will have not only to sustain the harsher environment, but also to 
 show high geometrical acceptance: for the future the material budget restrictions and the 
 geometrical limitations ask for geometry inefficiency to be below 2.5\%. 
 Hence the inactive areas of the future pixel sensor should be less than 450
  $\mu$m wide~\cite{IBLTDR}.

Different sensor options are being developed in parallel 
to address the challenges imposed by the
forseen luminosity upgrades. They include 
 diamond sensors~\cite{Diamond} and 3D-sensors, with implants
going through the silicon bulk~\cite{3D}. Optimizing the well-known
technology of planar silicon pixel sensors for the
ATLAS detector at an upgraded LHC accelerator is carried
out within the Planar Pixel Sensor project~\cite{PPS:proj}.

%% file: project.tex
\section{The project}
\label{sec:project}

Planar pixel sensors are the current technology for the ATLAS pixel system and a standard 
for tracking detectors in High Energy Physics. 
A lot of experience in designing, optimizing and producing planar sensors has been accumulated 
in the last 30 years; during these three decades planar sensors proved to be a reliable technology. 
There is still a lot of research on planar sensors and a lot of suppliers are on the market, 
assuring high quality and relatively low cost productions. 
Hence planar sensor are one good option for future tracking detectors. 
The ATLAS Upgrade Planar Pixel Sensor (PPS) R\&D Project is a collaboration of 17 groups 
and more 
than 80 scientists aiming to explore the suitability of planar pixel sensors for the ATLAS upgrade. 

Primary goal of the project is to demonstrate that planar sensors have the requested 
radiation hardness for the HL-LHC. 
To investigate this, several planar sensors technologies are under evaluation, 
including p-bulk (Section~\ref{sec:p_bulk}) and n-bulk (Section~\ref{sec:n_bulk}) option; 
regardless of the bulk, the group is looking only at sensors collecting electrons, based 
on their smaller trapping time compared to holes~\cite{Trapping}.

A reduction in sensor thickness (with respect to the 250 $\mu$m of the current ATLAS pixels) 
is investigated (Section~\ref{sec:thin}), to reduce the effects of trapping after irradiation 
and the material budget.

In case of a complete replacement of the Inner Detector a solution with more than 4 layers 
of pixels is possible. In this way an area of roughly 10 m$^2$ should be equipped with 
pixel modules. Hence it is crucial to minimize the costs of the future pixel system. 
 PPS  is evaluating several cost-reduction scenarios, including new bump-bonding 
technologies, given the fact that the bump-bonding cost dominated the cost of the current ATLAS pixel detector. The possibility to use p-bulk sensor is favorable 
in the view of  a cost reduction, having only one side patterned contrary to two-side patterned 
n-bulk sensors. 

The new sensors will have to assure a close to 100\% geometrical acceptance, so different 
options toward a reduction of the inactive area are pursued in the collaboration, such as slim edge 
geometries(Section~\ref{sec:slim}) and active edge detectors (Section~\ref{sec:active}).

Many of the new ideas for sensors are tested in advance by means of sensor simulations 
(Section~\ref{sec:simu}) thanks to a TCAD software. 

In Section~\ref{sec:outl} the plans of the PPS R\&D collaboration will be outlined, together with 
some general remarks on the work done so far.

%% file: p_bulk.tex
\section{P-bulk overview}
\label{sec:p_bulk}

In this Section a quick overview of the PPS activities on p-bulk sensors will be given.

Standard n-in-n sensors need guard-rings opposite to pixel side: the production requires 
to pattern both sides of the wafer. On the contrary, p-bulk does not need such a double 
processing, having only an homogeneous implantation on the back side.
This should allow for a significant cost reduction.
 Another major advantage of p-bulk over n-bulk is that there is no type inversion for irradiated 
material~\cite{moll-thesis}.

\begin{figure}[!htb]
\centering
\includegraphics[height=0.15\textheight]{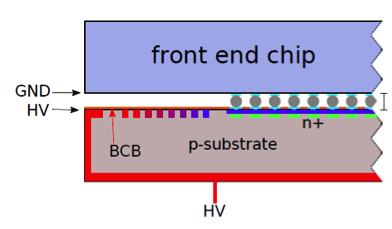}
\caption{\label{fig:n-in-p}P-bulk sensor sketch; potential of the different parts 
are indicated, while detectoro is biased.}
\end{figure}

A problem under investigation for the n-in-p detectors is the drop between the high voltage 
applied to the back side and the readout electronics potential (~0V) which occurs
solely on the sensor top side facing the readout electronics.
In figure~\ref{fig:n-in-p} a sketch of the p-bulk sensor geometry is displayed, 
highlighting the issue with the 
voltage drop on the sensor in a region close to the electronics.

Covering the sensor with a thin layer of BCB (Benzo Cyclo Butene) proved to prevent 
sparks up to 1000 V.

Several contributors are working with  p-bulk sensors in the PPS R\&D group, namely 
 CiS Forschungsinstitut f\"ur Mikrosensorik und Photovoltaik GmbH, 
Micron Semiconductor Ltd., 
Hamamatsu Photonics K.K. (HPK) 
and Max Planck Institut  f\"ur Physik - Max Planck Institut Halbleiterlabor (MPI-HLL). 
 A variety of sensor thicknesses is currently investigated by the different contributors.
In table~\ref{tab:n-in-p} a summary is given.  

\begin{table}[!htb]
\begin{center}
\begin{tabular}{|l|c|}
\hline
Suppliers & Thickness ($\mu m$) \\
\hline 
 CiS & 285/200/150 \\
\hline 
Micron & 300/150 \\
 \hline
 HPK & 320/150 \\
 \hline
 MPI-HLL & 150/75 \\
 \hline
\end{tabular}
\end{center}
\caption{\label{tab:n-in-p}P-bulk sensor suppliers and relative wafer thicknesses}
\end{table}

More details are given in~\cite{Anna}. 

%% file: n_bulk.tex
\section{N-bulk sensors: results for irradiated devices}
\label{sec:n_bulk}

To assess the radiation hardness of n-bulk sensors, irradiation campaigns have been 
carried out and crucial observables, such as the charge colleced per impinging particle and 
hit efficiency, were studied as a function of the delivered dose. In the following details 
on the studies performed on irradiated n-in-n sensors will be given.

\begin{figure}[!htb]
\begin{center}
\includegraphics[height=0.175\textheight]{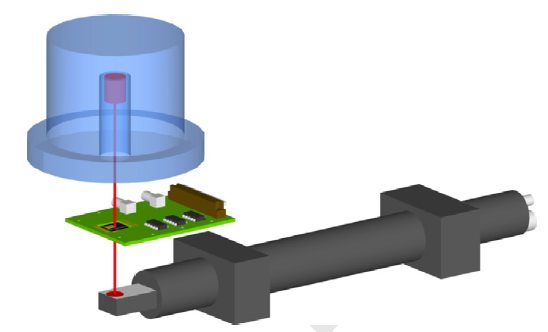}
\includegraphics[height=0.175\textheight]{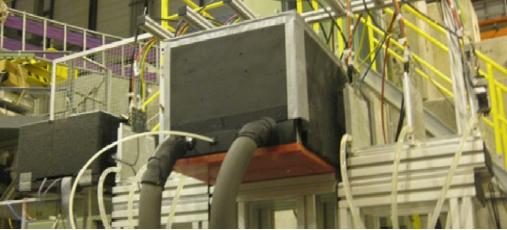}
\end{center}
\caption{\label{fig:n-tools}left: Schematics of the $^{90}$Sr setup, with source, collimator, module 
and scintillator detector; right: the CERN SPS testbeam setup. Pions were incoming from the 
right, traversing the first arm of the telescope, then the black box with the DUTs, }
\end{figure}

The starting point for the PPS R\&D project are the ATLAS single chip pixel 
modules~\cite{AtlasPixels}. 
Planar n-in-n ATLAS sensors, already assembled with a FE-I3 chip in a single chip module  (SCM) 
were irradiated with neutrons at the TRIGA nuclear reactor~\cite{Triga} of the Jozef Stefan
Institute in Ljubljana. 
Modules were irradiated with fluences up to $2\times 10^{16} n_{eq}/cm^{2}$ and then 
characterized with radioactive sources and beams~\cite{nTUD}. 
In figure~\ref{fig:n-tools} the two experimental setups are shown.

 As displayed on the left, 
the $^{90}$Sr source was enclosed by a collimator.
The module was cooled  with dry ice 
and operated at temperatures below -50$^{\circ}$ C. 
A trigger scintillator was mounted below the SCM. 

The SCMs were also operated in testbeams at the
 CERN SPS beamline (see figure~\ref{fig:n-tools}, right) in July and October 2010.
 Thanks to the testbeam, it was possible
  to compare the energy depositions of high-energy pions with
those of beta-electrons and to be able to observe the effect of
different incidence angles after high fluences. The testbeam was
conducted with the EUDET beam telescope~\cite{EUDET} which uses 
six MIMOSA26 chips~\cite{mimosa26} as telescope planes.
The  six telescope planes were defining the tracks 
  and up to six devices under test (DUTs) were put between the first and second telescope arm. 
  The irradiated modules 
   were cooled with dry ice; the operation 
   temperature was kept between -35 and -15$^\circ$ C.

\begin{figure}[!htb]
\begin{center}
\includegraphics[height=0.20\textheight]{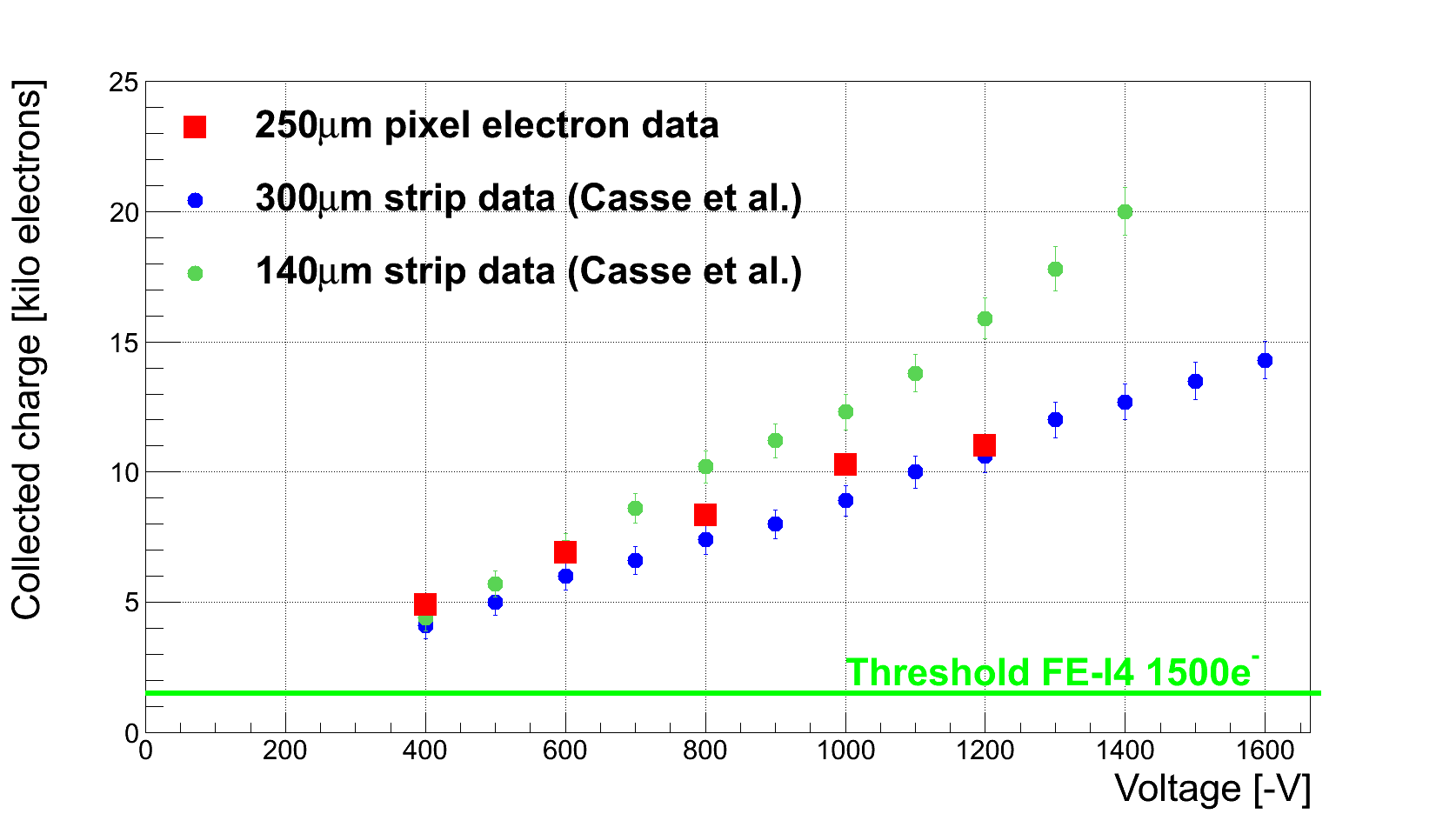}
\includegraphics[height=0.20\textheight]{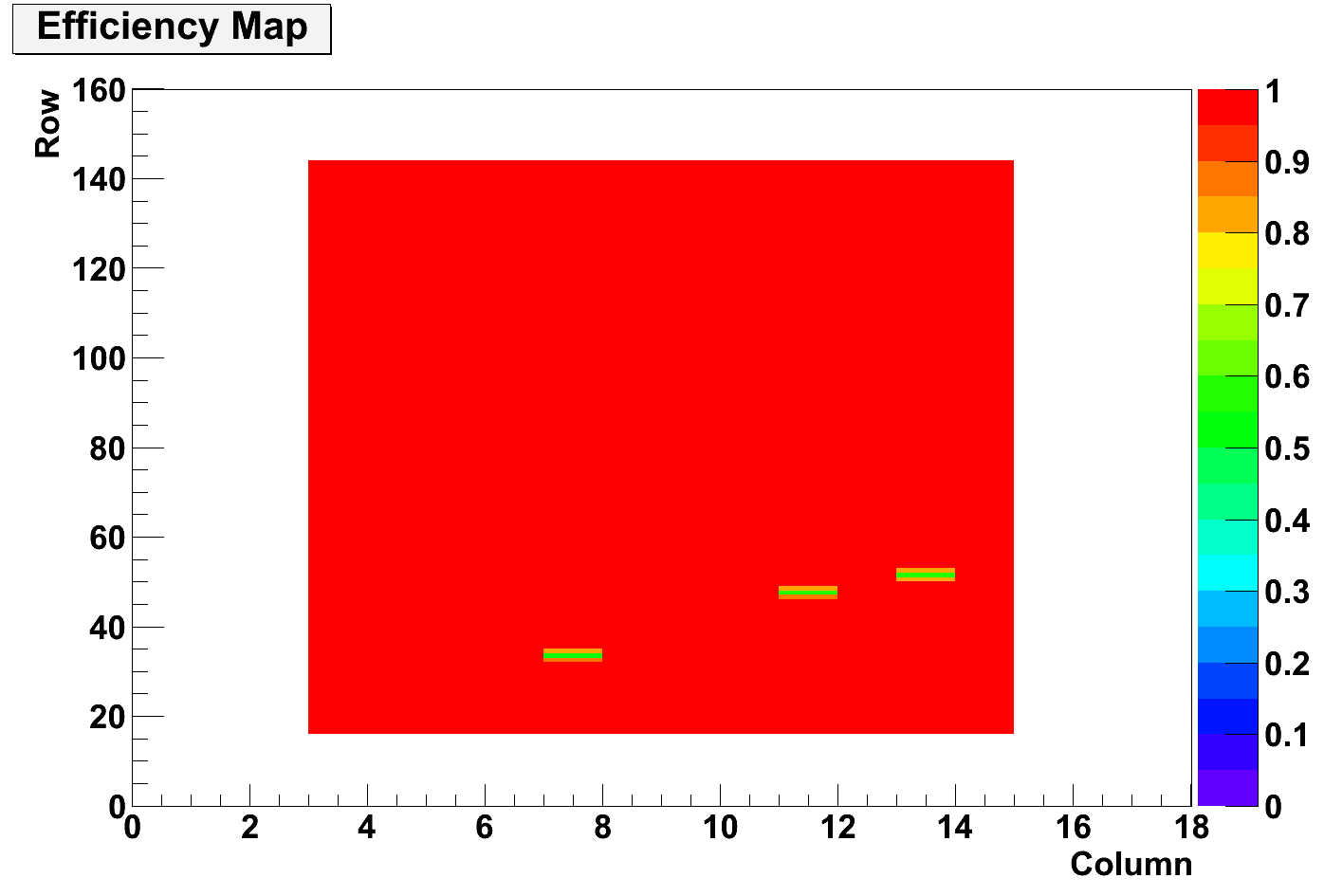}
\end{center}
\caption{\label{fig:5e15}left: Collected charge as a function of the bias for sensors irradiated 
with fluence of $5\times 10^{15} n_{eq}/cm^{2}$; right: Hit efficiency map for the same assembly}
\end{figure}

In figure~\ref{fig:5e15}, on the left, a study of the collected charge vs bias voltage
 for a module irradiated with a fluence of $5\times 10^{15} n_{eq}/cm^{2}$ is shown\footnote{as 
 a comparison data from~\cite{Casse} are reported too.} (only data collected with the 
 $^{90}$Sr are shown).  In the figure the expected threshold for the FE-I4 
 chip is shows too. A signal of about 10 ke$^-$ is observed at 1000 V. This is very 
 promising for the foreseen ATLAS $4^{\rm th}$ pixel layer (``IBL'')~\cite{IBL}, and 
  the outer pixel layers at HL-LHC. 

In figure~\ref{fig:5e15}, on the right, the hit-efficiency map for the module irradiated with a 
fluence of $5\times 10^{15} n_{eq}/cm^{2}$, as measured at the testbeam for particles 
at normal incidence. The SCM was biased at 1000 V and the hit-efficiency was 99.6\%. 

Hit efficiency was studied as a function of the bias voltage, for the 
module irradiated with $5\times 10^{15} n_{eq}/cm^{2}$. 
Results are in table~\ref{tab:n-eff}.

\begin{table}[!htb]
\begin{center}
\begin{tabular}{|c|c|c|}
\hline
Bias voltage (V) & Hit efficiency (\%) \\
\hline
350 & 93.2 \\
\hline
500 & 97.3 \\
\hline
1000 & 99.6 \\
\hline
\end{tabular}
\end{center}
\caption{\label{tab:n-eff}Hit efficiency of an irradiated (fluence = $5\times 10^{15} n_{eq}/cm^{2}$) 
FEI3 n-in-n module at different bias voltages.}
\end{table}

The results for n-in-n devices indicate that for the second layer at HL-LHC, current pixel  
system is able to collect sensible amount of charge with efficiency close to 1.

\begin{figure}[!htb]
\begin{center}
\includegraphics[height=0.25\textheight]{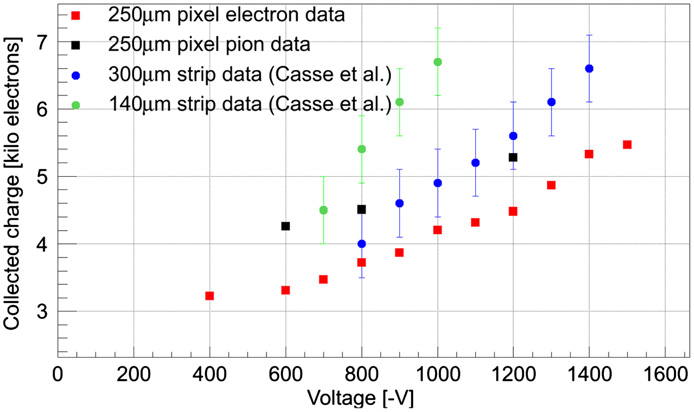}
\end{center}
\caption{\label{fig:2e16}Collected charge as a function of the bias for sensors irradiated 
with fluence of $2\times 10^{16} n_{eq}/cm^{2}$; for comparison, strip-data
 from~\cite{Casse2e16} are added.}
\end{figure}

The module irradiated with a fluence of $2\times 10^{16} n_{eq}/cm^{2}$ was able to collect 
roughly 5500 electrons at 1500 V; the result is presented in figure~\ref{fig:2e16}.
This remarkable result shows the possibility to still collect charge with 
planar detectors at the highest expected fluences of HL-LHC.

The results for the collected charge by the n-in-n irradiated module are summarized in 
table~\ref{tab:n-in-n-fluence}.

\begin{table}[!htb]
\begin{center}
\begin{tabular}{|c|c|c|}
\hline
Fluence ($10^{15} n_{eq}/cm^{2}$) & Bias voltage (V)  & Collected charge (10 ke$^-$) \\
\hline 
 1 & 1000 & 18 \\
\hline 
5 & 1000 & 10 \\
 \hline
20 & 1500 & 5.5 \\
 \hline
\end{tabular}
\end{center}
\caption{\label{tab:n-in-n-fluence}Charge collected by neutron irradiated n-bulk modules.}
\end{table}

%% file: thin.tex
\section{Thin production}
\label{sec:thin}

While thin planar sensors yield lower - nonetheless sufficient
- signals before irradiation, they offer several potential
advantages compared to sensors with standard thicknesses.
A reduction of the radiation length in silicon leads to less
multiple scattering allowing for improvements in the tracking
resolution. Furthermore, for the same voltage the electric field
in thin sensors is higher. This seems to increase the signal size
after irradiations by charge multiplication effects~\cite{Qmult}.

Recent results~\cite{MultCasse} indicate that with a 140 $\mu$m thick sensor irradiated with 
$5\times 10^{15} n_{eq}/cm^{2}$ it is possible to collect 12 thousand electrons at 1000 V, 
roughly 2000 electrons more than a standard 300 $\mu$m thick planar sensor.

A dedicated n-wafer production with thicknesses ranging from standard 250 $\mu$m down to 
150 $\mu$m has been competed at CiS. 
First modules have been assembled and are currently being irradiated.
They will be used for detailed studies of charge collection 
and charge amplification. 

%% file: slim.tex
\section{Slim edge}
\label{sec:slim}

In the present pixel sensor the distance between the active
pixel area and the cutting edge is  1100~$\mu$m. This was done to prevent the 
depletion area to reach the cut-edge.

The geometry of the future ATLAS pixel detector foresees no shingling in $z$ direction, 
and flat, single-sided staves for the inner radii, hence 
a reduction in sensor dead-area is required.

\begin{figure}[!htb]
\begin{center}
\includegraphics[height=0.20\textheight]{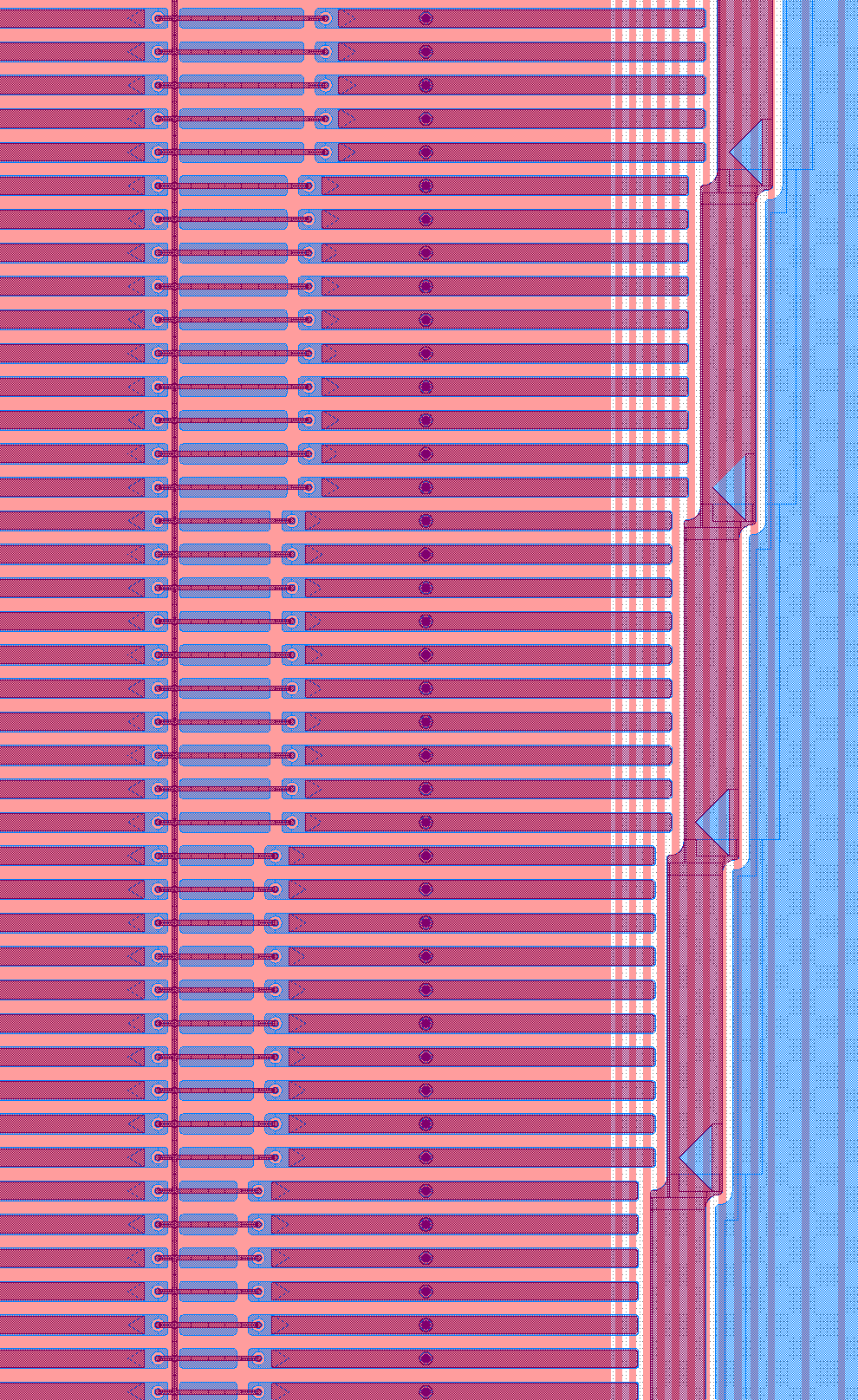}
\includegraphics[height=0.20\textheight]{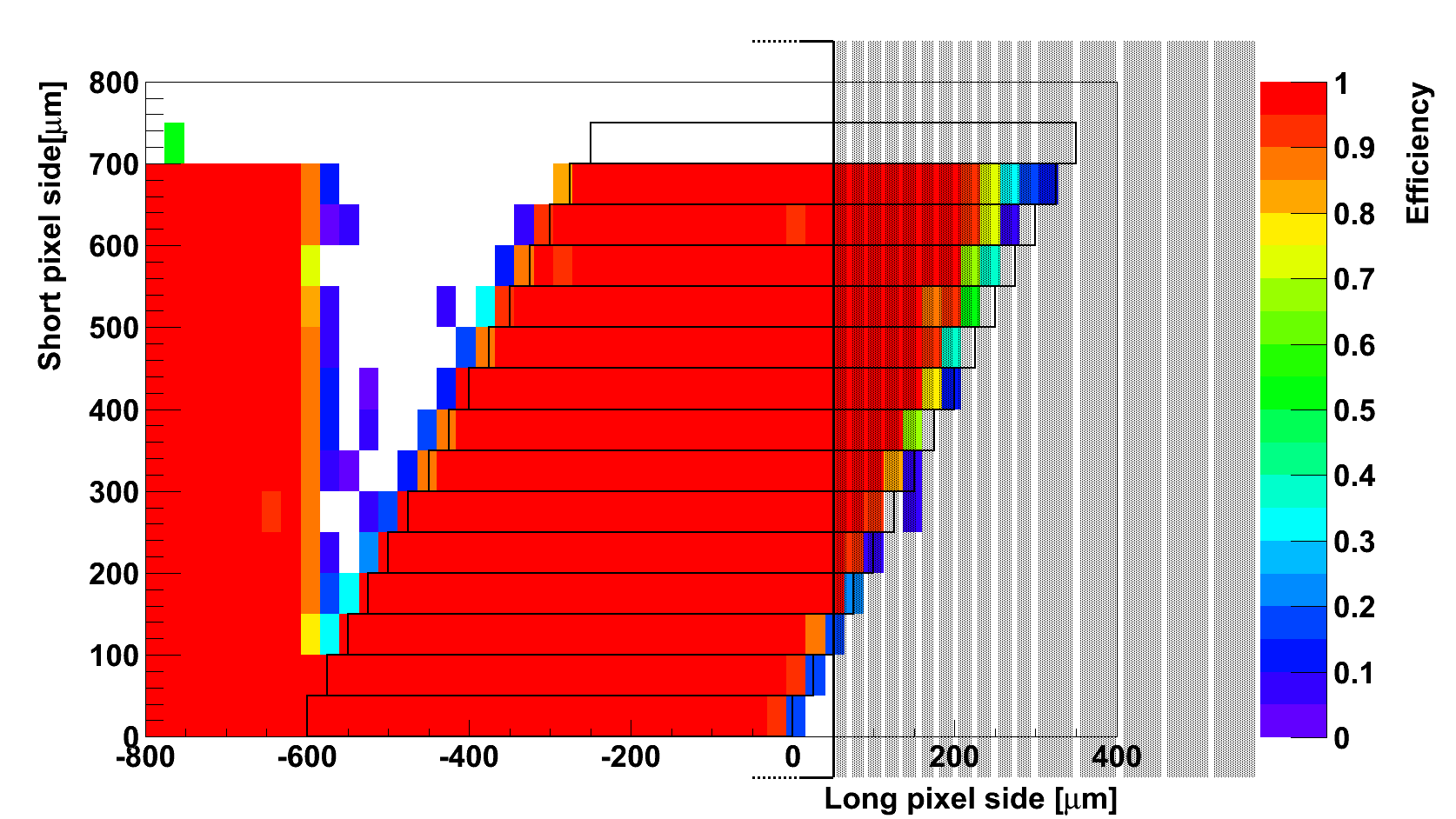}
\end{center}
\caption{\label{fig:slim_eff}left: ``stepwise pixel'' detector design; 
right: hit-efficiency map for a non-irradiated  `stepwise pixel'' detector, 
as observed at the testbeam. Grey part represents the guard-ring area.}
\end{figure}

 PPS  studied an n-in-n pixel sensor in which the pixels close to the edge are longer 
than the standard ones, partially overlapping with the guard-ring region 
(``slim edge'' design). 
A dedicated structure to this goal was prepared, with pixels shifted in a step-wise way
 (``stepwise pixel'') in order to see 
which was the minimum inactive area with still excellent hit-efficiency.
In figure~\ref{fig:slim_eff}, on the left, the design of this detector is illustrated; dark red horizontal 
rectangles correspond to pixel implants, while  light red vertical strips (at the right edge)
 represent guard-rings.

In figure~\ref{fig:slim_eff}, on the right, 
a hit-efficiency map (as observed at the testbeam mentioned in 
section~\ref{sec:n_bulk}) for a non-irradiated ``stepwise'' pixel detector is presented.
The p+ implant on the back side ends at 0 and  the grey part represents the guard-ring area (on the opposite side of the pixels), which is 
part of the ``inactive'' area in the standard design.
 It is clear that pixels are almost 100\% 
efficient up to 200 $\mu$m from the p+ implant edge, reducing in this way the inactive area to
 250 $\mu$m only.

%% file: active.tex
\section{Active edge}
\label{sec:active}

As already discussed 
detectors made using planar technology have an insensitive
region around their edges to prevent the depletion region to reach 
the  nonpassivated saw cuts at the detector edges and to allocate
enough space for the guard rings~\cite{3dKenney}.
Nonetheless reducing the inactive area between the edge pixels and the cut-edge is essential to 
 maintain a high geometrical acceptance.
 
To control the potential drop along the cut edge, and then reduce the 
inactive area several ``active edge'' approaches are under study.
 
One possibility is to terminate the sensor with heavily doped trenches, where the insensitive 
edge region can be reduced to a few $\mu$m; this technique was originally adopted by FBK for 
 3D sensors~\cite{activeFBK}. A trench is dug by Deep Reactive Ion Etching (DRIE), then it is doped 
 and filled with polysilcon. 
 This process allows the production of particularly thin sensors: thus it is very promising 
 for inner HL-LHC pixel layers.
 Simulations carried inside the PPS collaboration indicate the possibility to 
 realize an n-in-p detector with reduced dead area (few tens of $\mu$m) and good stability, even 
 after large irradiation.

 Another approach is to passivate the cutting edge 
 by atomic layer deposition (ALD)~\cite{activeScipp}. 
A different approach for n-bulk and p-bulk sensor is used. In the former case a positively charged 
layer is envisaged, so a low-temperature oxide growth is performed; 
in the latter case a negatively charged 
layer is needed, and so an ALD of Al$_2$O$_3$ is done. 

\begin{figure}[!htb]
\begin{center}
\includegraphics[height=0.23\textheight]{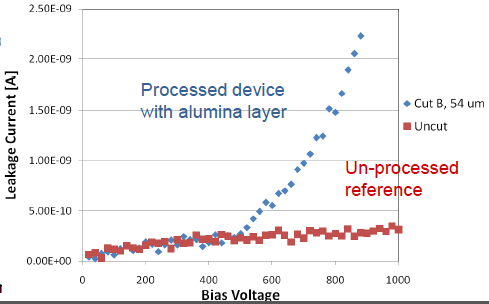}
\caption{\label{fig:scippDiode} IV curve for a n-in-p diode with passivated trench; red points 
correspond to an un-cut un-processed diode.}
\end{center}
\end{figure}

First test on diodes with only one guard ring~\cite{Unno07} indicates no breakdown up to 500 V 
after the cut  and stable current below 50 pA~\cite{activeScipp}; IV curve for this device is shown 
in figure~\ref{fig:scippDiode}.

The passivation of the trench is performed after cutting the sensor from the wafers (several 
cutting techniques are considered~\cite{Sadro2009}). This feature makes this approach interesting 
for outer layers as it is possible to use with sensors from any vendor and type.

%% file: simu.tex
\section{Advanced simulations}
\label{sec:simu}

In the PPS project several simulation activities are ongoing,
trying to identify possible ways to optimize the layout and
improve the sensor performance. 

The program 
Silvaco Technology Computer Aided Design (TCAD)~\cite{Silvaco} is used for 
simulating lithography and
implantation parameters to model and evaluate various semiconductor
designs and processes, helping in reducing the number of wafer submission 
by identifying early problematic layout configurations or potentially 
non-optimal process parameters.

The electrical characteristics of semiconductor devices are simulated starting from 
the definition of the processes to form the diode junction, such as oxidation of the 
bare bulk material, insulation implantation, oxide etching, acceptor/donor implantation 
for electrical contacts definition and so on. 
Once the structure of the semiconductor device is defined it is possible to simulate 
its behavior under several realistic situations, such as applied bias voltages, 
radiation-induced damages in the bulk and on the surface, illumination by laser, etc. 
All the relevant observables, such as currents, capacitance, distribution of 
electric potential and field, of free carriers and of space charge can be studied as a function 
of the model parameters.

The PPS project studied by means of Silvaco-TCAD simulation the behavior of the detector 
after irradiation, in relation to different designs for the multi-guardrings 
structures~\cite{CalderiniHiroshima, Benoit:2009zz}.
The radiation damage was modeled by localized defects in the band gap of silicon. 
Several models exist; among them the one based on the RD50 collaboration 
results~\cite{rd50-2007} 
and work of several groups~\cite{Moscatelli-2002, Moscatelli-2004, Moscatelli-2006}.

\begin{figure}[!htb]
\begin{center}
\includegraphics[height=0.18\textheight]{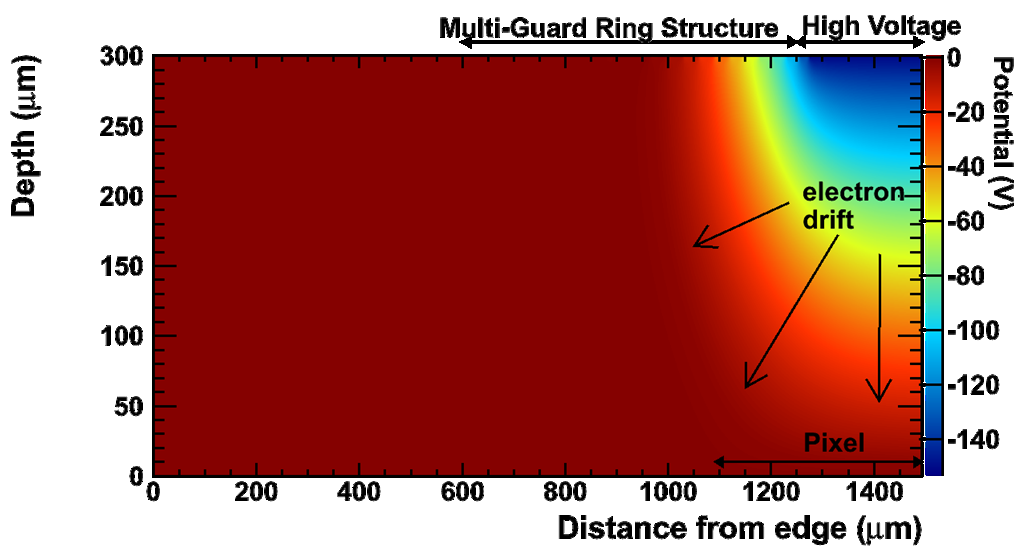}
\includegraphics[height=0.18\textheight]{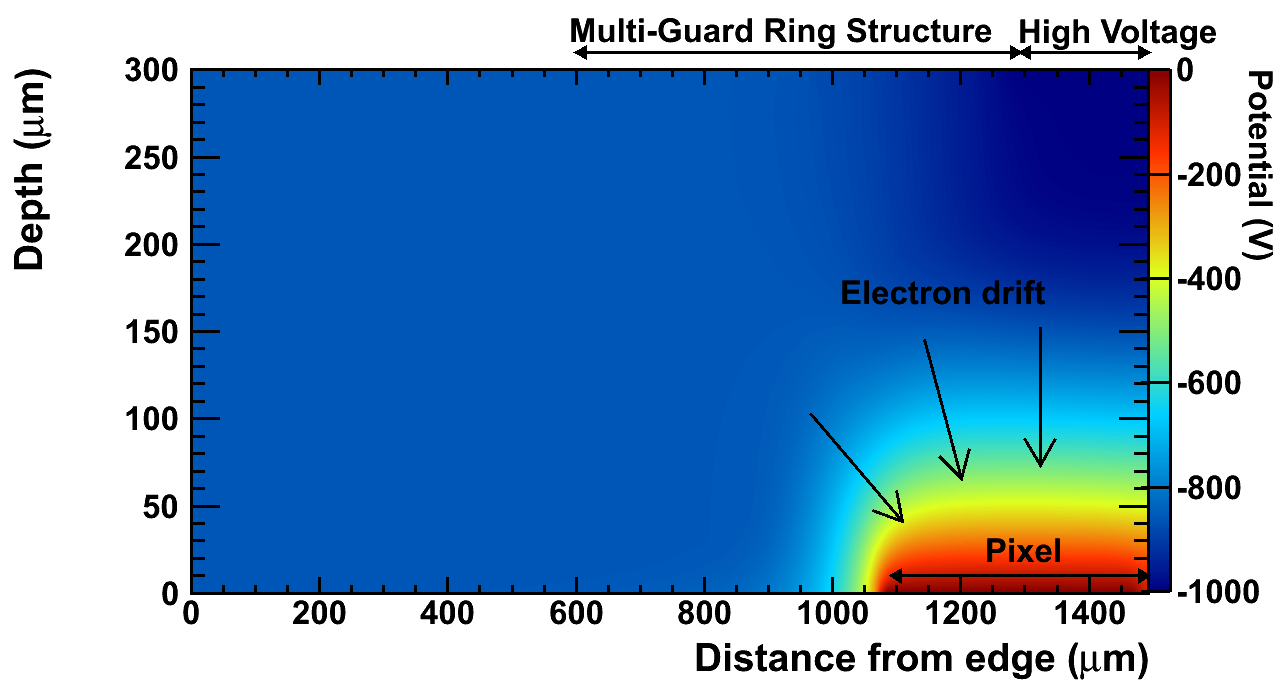}
\caption{\label{fig:simu}Potential maps for simulated n-in-n slim-edge pixels. 
Left: non irradiated device. Right:  $1\times 10^{15} n_{eq}/cm^{2}$ irradiated one.}
\end{center}
\end{figure}

As an example of simulations application we show on figure~\ref{fig:simu} the results 
for slim edge design (see section~\ref{sec:slim}); on both pictures the potential map is 
shown, before (left) and after (right) irradiation (fluence of $1\times 10^{15} n_{eq}/cm^{2}$). 
We can observe that after irradiation the electric field lines are less distorted than before; 
this is a nice prediction reassuring us on the viability of the slim edge design even after 
large fluences.

%% file: outl.tex
\section{Conclusions \& Outlook}
\label{sec:outl}

The ATLAS PPS R\&D Project is trying to asses the radiation hardness of the planar pixel 
sensors, in view of the luminosity upgrade foreseen for the LHC (the so called HL-LHC era). 

The PPS group is investigating both n- and p-bulk sensors. 
First sensor prototypes are been evaluated with radioactive source and on beams. 
Results for n-in-n devices indicate the viability of the technology (large signal collected for MIPs 
with close to 1 efficiency) even after irradiation at fluences up to  $5\times 10^{15} n_{eq}/cm^{2}$ 
(the expected fluence for outer pixel layers at HL-LHC). 
Pixels shown to collect sizable amount of charge even at $2\times 10^{16} n_{eq}/cm^{2}$, 
the expected fluence for the innermost pixel layers at HL-LHC.

The group is committed into inactive area reduction, to cope with the geometrical restrictions for  
future ATLAS tracker, keeping geometrical acceptance close to 1. 
Several approaches are investigated: a ``slim edge'' design, by pushing 
pixels beyond guard-ring boundary, and ``active edge'' techniques, which deal with the treatment 
of the trench at edge. 

Several studies are stimulated first, and optimized then, by dedicate simulations of 
the proposed sensors. 

New sensor designs are under way, which will be throughly tested with radioactive source and 
in beam, after irradiation; the final goal will be to prove that planar pixels can be the 
 technology for the ATLAS tracker in the high luminosity LHC era, too.